\documentclass[twocolumn,secnumarabic,amssymb, nobibnotes, aps, prl, superscriptaddress, nobalancelastpage,longbibliography]{revtex4-1}

 \usepackage{graphicx}
\usepackage{bm}
\usepackage{amsmath} \usepackage{braket} \usepackage{epsfig}
\usepackage{tensor}
\usepackage{CJKutf8}
\usepackage[version=4]{mhchem}
\usepackage{titlesec}
\usepackage{ifthen}
\usepackage{sidecap}
\usepackage{listings}
\usepackage[para,online,flushleft]{threeparttablex}

\usepackage{xr-hyper}
\usepackage{hyperref}
\hypersetup{breaklinks=true,colorlinks=true,linkcolor=blue,citecolor=blue,filecolor=magenta,urlcolor=cyan}

\usepackage[all]{hypcap}

\usepackage{xcolor}
\definecolor{pastelgray}{rgb}{0.81, 0.81, 0.77}
\definecolor{beaublue}{rgb}{0.9, 0.9, 0.93}

\makeatletter
\def\@bibdataout@aps{%
\immediate\write\@bibdataout{%
@CONTROL{%
apsrev41Control%
\longbibliography@sw{%
    ,author="08",editor="1",pages="1",title="0",year="1"%
    }{%
    ,author="08",editor="1",pages="1",title="",year="1"%
    }%
  }%
}%
\if@filesw \immediate \write \@auxout {\string \citation {apsrev41Control}}\fi
}
\makeatother

\renewcommand{\vec}[1]{\mbox{\boldmath $#1$}}

\begin{document}
\begin{CJK*}{UTF8}{gbsn}

\title{Fermion Pair Dynamics in Open Quantum Systems}

\author{S.M. Wang (王思敏)}
\affiliation{FRIB Laboratory, Michigan State University, East Lansing, Michigan 48824, USA}
\affiliation{School of Physics, and State Key Laboratory of Nuclear Physics and Technology, Peking University, Beijing 100871, China}

\author{W. Nazarewicz}
\affiliation{Department of Physics and Astronomy and FRIB Laboratory, Michigan State University, East Lansing, Michigan 48824, USA}

\date{\today}

\begin{abstract}
Three-body decay is a rare decay mode observed in a handful of unbound rare isotopes. The  angular and energy correlations between emitted nucleons are of particular interest, as they provide invaluable information on the interplay between structure and reaction aspects of the nuclear open quantum system. To study  the mechanism of two-nucleon emission, we developed a  time-dependent approach that allows us to probe emitted nucleons at long times and large distances. We successfully benchmarked the new method against 
the  Green's function approach and applied it to low-energy two-proton and two-neutron decays. In particular, 
we studied the interplay between initial-state nucleon-nucleon correlations and final-state interaction. We demonstrated that the  time evolution of the two-nucleon wave function is strongly impacted by the diproton/dineutron dynamics and that
the correlations  between emitted nucleons provide invaluable information on the dinucleon structure in the  initial-state.

\end{abstract}
\maketitle

\end{CJK*}

{\it Introduction}.---The nucleus is a unique laboratory of quantum many-body physics. Its fermionic building blocks, the positively-charged proton and the neutral neutron, are almost identical in all  aspects, except for the electric charge. This is a consequence of the isospin symmetry \cite{Wilkinson1970,Warner2006}. The symmetry is weakly broken in atomic nuclei, primarily by the long-range Coulomb interaction. The spectroscopy of mirror nuclei, which have their proton number and neutron number exchanged, offers many  examples of isospin-symmetry violation. Important, complementary information on the interplay between the nuclear and electromagnetic force comes from  decay studies, including  two-proton ($2p$) and two-neutron ($2n$) radioactivity \cite{Goldansky1960,Blank2008,Pfutzner2012}. In this respect,  particularly valuable is the information  on the angular and energy correlations between emitted nucleons, which reflect the interplay between structure and reaction aspects of the nuclear open quantum system \cite{Michel2010a}. A more general perspective on two-nucleon decays is offered by its relevance to  quantum entanglement \cite{Bertulani2003}.

A handful of ground state (g.s.) 2$p$ emitters have been discovered  \cite{Blank2008,Pfutzner2012}. In the case of the   g.s.  2$n$ decay, the only candidate identified so far  is
 the threshold resonance in $^{26}$O  \cite{Kohley2015,Kondo2016,Grigorenko2013}.
Theoretically, the mechanism of two-nucleon emission  is not fully understood, including the role of the low-lying particle continuum, and the transition from the initial state  strongly impacted by the  nuclear medium to the scattering domain governed by final-state interaction \cite{Watson1952,Migdal1955,Alt2003}. In this Letter we study the  two-nucleon decay process in a time-dependent framework to address this question.

At the initial stage of the two-particle decay process, the emitted nucleons are highly correlated due to the presence of nucleonic pairing.
In particular, due to the low-lying particle continuum, the correlated nucleonic pairs in the surface region of superfluid weakly bound nuclei (often referred to as di-nucleons) are believed to be spatially compact \cite{Pillet2007,Hagino2005,Hagino2007,Matsuo2012,Hagino2014}. 

The majority of theoretical models of two-nucleon radioactivity -- such as the WKB approach \cite{Goncalves2017,Nazarewicz1996}, $R$-matrix theory \cite{Barker2003,Brown2003}, and the three-body reaction model of Refs.~\cite{Grigorenko2009,Grigorenko2017} -- treat internal and asymptotic regions separately. Except for the three-body models,  these approaches are based on the assumption that the inner nuclear structure could be preserved during the course of tunneling. Since the dinucleon is  unstable, this
supposition is unlikely to hold. To describe two-nucleon decay  comprehensively, in our previous work \cite{Wang2017,Wang2018}, we introduced the Gamow coupled-channel (GCC) method, which is capable of  describing structure and decay aspects of  unbound three-body systems within one   framework. However, to model angular and energy correlations of emitted particles, precise three-body solutions at very large distances are required, and this poses a formidable challenge, especially in the presence of the long-range Coulomb interaction.

An alternative strategy of tackling the decay process is a time-dependent formalism, which allows to address a broad range of questions -- such as configuration evolution \cite{Volya2014}, decaying rate \cite{Peshkin2014}, and fission \cite{Bender2020}  -- in a  precise, numerically stable, and transparent way. In the case of two-nucleon decay, the measured inter-particle correlations can be interpreted in terms of solutions propagated for long times. An approximate treatment of $2p$ emission was proposed in Ref.\,\cite{Bertulani2008}, in which the center-of-mass motion for the two-protons was described classically. In Refs.\,\cite{Oishi2014,Oishi2017}, an early stage of the two-proton emission from the g.s. of $^{6}$Be was investigated using a time-dependent method. However,  their method is not  able to capture the asymptotic dynamics and resulting proton-proton ($pp$) correlations.
To this end, we have developed a realistic time-dependent approach to investigate the general properties and unique features of $2p$ and $2n$ radioactivity. Our method provides precise solutions at very large distances (over 500\,fm) and long times (up to 30\,pm/$c$) that  allows  analysis of asymptotic observables in order to compare them with  measurements.

\begin{figure}[!htb]
\includegraphics[width=1.0\columnwidth]{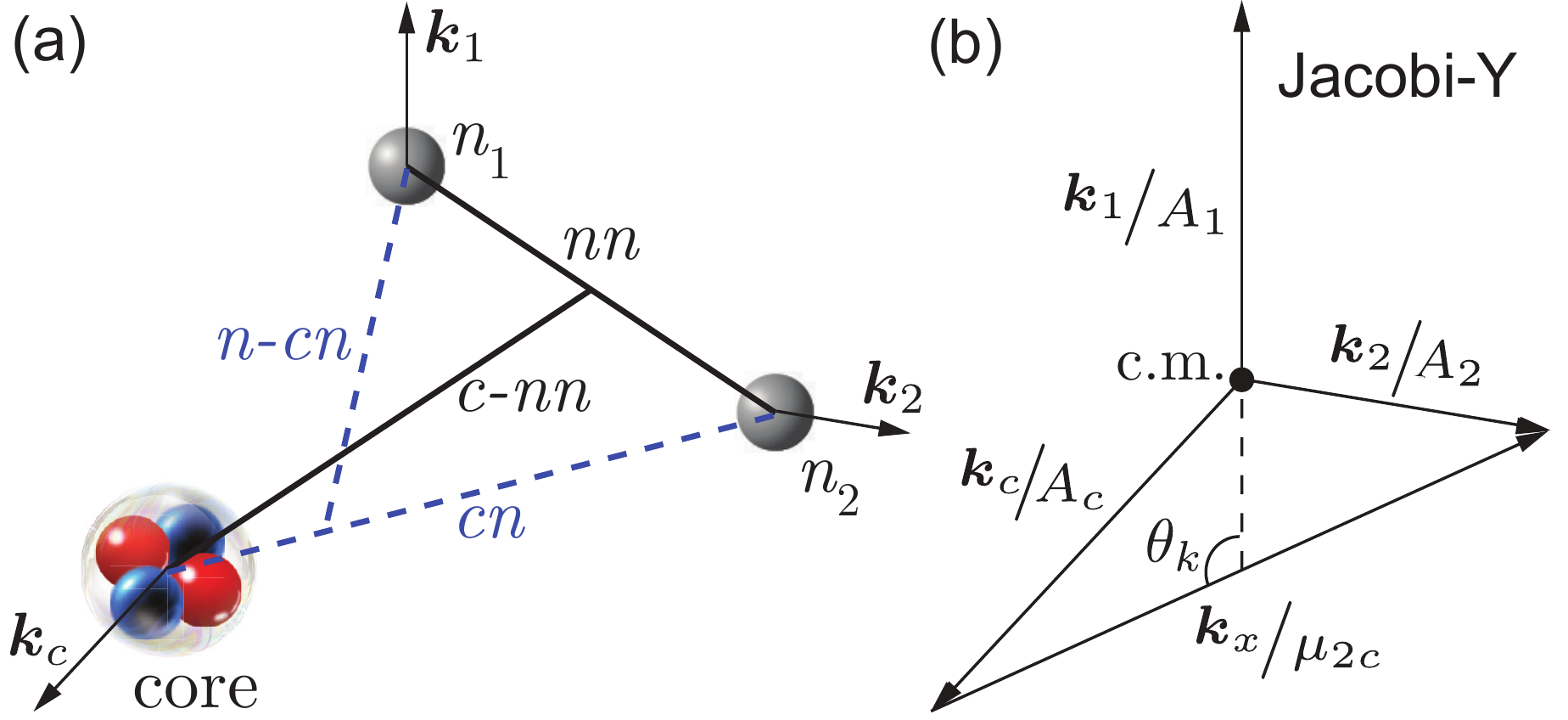}
\caption{(a) Jacobi T (solid lines) and Y (dashed lines) coordinates for  a core+nucleon+nucleon system and (b) the corresponding momentum scheme. $A$ is the mass number, and $\mu_{ij}$ is the reduced mass of clusters $i$ and $j$. $k_1$, $k_2$ and $k_c$ are the momenta of  the nucleons $n_1$ and $n_2$ and the core $c$, respectively in the center-of-mass (c.m.) coordinate frame.}\label{Jacobi_coordinates}
\end{figure}

{\it Theoretical approach}.---A two-nucleon emitter can be viewed as a three-body system: a core ($c$) representing the daughter nucleus and two emitted nucleons ($n_1$ and $n_2$). 
The $i$-th cluster ($i=c,n_1,n_2$)
has the position vector  $\vec{r}_i$ and linear momentum $\vec{k}_i$. The corresponding Hamiltonian can be written as:
\begin{equation}\label{Hcnn}
	\hat{H} = \sum_i\frac{ \hat{\vec{p}}^2_i}{2 m_i} +\sum_{i>j} \hat{V}_{ij}(\vec{r}_{ij}) -\hat{ T}_{\rm c.m.},
\end{equation}
where $\hat{V}_{ij}$ represents the two-body interaction between the constituent clusters and $\hat{T}_{\rm c.m.}$ stands for the center-of-mass term.

In order to describe three-body asymptotics and to eliminate the spurious center-of-mass (c.m.) motion, it is convenient to build the total wave function $\Psi^{J\pi}$ in the relative (Jacobi) coordinates (see Fig.\,\ref{Jacobi_coordinates} and Supplemental Material (SM) \cite{SM}) using the hyperspherical-harmonics expansion. To obtain the initial resonance state $\Psi_0^{J\pi}$ at $t=0$, we take advantage of the GCC framework and extend the Schr\"odinger equation into the complex-momentum $\tilde{k}$ space by utilizing the Berggren  expansion \cite{Berggren1968,Michel2009,Wang2017}.  The Berggren basis includes resonant and scattering states; hence, it effectively allows the treatment of nuclear structure and reactions. Such a Jacobi-Berggren  approach describes decay properties of outgoing valence nucleons and prevents the reflection of the wave function at the boundary. This offers a significant advantage over the use   of the cluster-orbital-shell-model coordinates \cite{Suzuki1988}  used in the standard Gamow Shell Model applications  \cite{Michel2009}, and the use of absorbing boundary condition as in Refs.~\cite{Oishi2014,Oishi2017,Oishi2018}. The Pauli-forbidden states occupied by the core nucleons are eliminated through the supersymmetric transformation method \cite{Sparenberg1997}, which preserves the phase and spectral equivalence.

The complex-momentum state $\Psi_0 ^{J\pi}$ obtained with the GCC method can be decomposed into real-momentum scattering states using the Fourier–Bessel series expansion in the real-energy Hilbert space \cite{Baz1969}. The resulting wave packet is propagated by the time evolution operator through the Chebyshev expansion \cite{Ikegami2002,Volya2009}:
\begin{equation}\label{time_propagator}
	e^{-i \frac{\hat{H}}{\hbar} t}=\sum_{n=0}^{\infty}(-i)^{n}\left(2-\delta_{n 0}\right) J_{n}(t) T_{n}(\hat{H}/\hbar),
\end{equation}
where $J_{n}$ are the Bessel functions of the first kind and $T_{n}$ are  the Chebyshev polynomials. We limit the time evolution to the real momentum space; this restores the Hermitian property of the Hamiltonian matrix and guarantees the conservation of the total density. Since our realization of the time-dependent approach is based on the integral equation, and since the Chebyshev expansion has a good convergence rate, the numerical precision can be well controlled to maintain high accuracy \cite{Loh2001,Volya2009}. Moreover, since the evolving wave packet has an implicit cutoff at large distances, the divergence of the Coulomb interaction in the momentum space is avoided. In practice, we only consider the interactions inside the sphere of radius 500\,fm, but the wave function is still defined in the momentum space beyond this cutoff. In this way, the unwanted reflection at the boundary is avoided.

{\it Hamiltonian and model parameters}.---
The three-body configurations in the Jacobi coordinates are labeled by  quantum numbers  $(K,\ell_x,\ell_y,S)$, where $K$ is the hyperspherical quantum number,  $\ell_x$ is the orbital angular momentum of the proton (neutron) pair with respect to their center of mass,  $\ell_y$ is the pair's orbital angular momentum with respect to the core, and $S$ is the total intrinsic-spin of the emitted nucleons. The calculations were carried out in a model space defined by $\max(\ell_{x}, \ell_{y})\le 7$ and for a maximal hyperspherical quantum number $K_{\rm max} = 50$. In the hyperradial part, we used the Berggren basis for the $K \le 10$ channels and the harmonic oscillator basis with the oscillator length of 1.75\,fm and $N_{\rm max} = 20$ for the remaining channels. For the GCC calculation of the initial state, the  complex-momentum contour defining the Berggren basis is given  by the path: $\tilde{k} = 0 \rightarrow 0.3-0.15i \rightarrow 0.5-0.12i  \rightarrow 1 \rightarrow 2 \rightarrow 4$ (all in fm$^{-1}$). As a result, the  initial state used  in our time-dependent calculations  is a resonance with a complex energy. The imaginary part of the energy, related to the decay width, represents the energy uncertainty of the wave packet.
For the time-dependent calculation, the inner part ($<15$\,fm) of the initial state is expanded and propagated with a real-momentum contour, which is $k = 0 \rightarrow 0.25 \rightarrow 0.5  \rightarrow 1 \rightarrow 2 \rightarrow 4$ (all in fm$^{-1}$). Each segment is discretized with 100  scattering states.

In this work, we consider the $2p$/$2n$ decays from a mirror pair $^6_4$Be$_2-^6_2$He$_4$.
Our aim is not to provide a detailed description of  the actual experimental data (for this, one would need a fine-tuned Hamiltonian) but rather to demonstrate the high precision of time-dependent solutions and explore the unique generic features of two-nucleon decays.  The interaction between the valence nucleons is represented by the finite-range Minnesota force with the original parameters of Ref.\,\cite{Thompson1977}, which is supplemented by the two-body Coulomb force for protons. The effective core-nucleon interaction has been taken in the form of a Woods-Saxon potential (with spin-orbit term) and a one-body Coulomb interaction. Except for the potential depth $V_0$, the parameters of the Woods-Saxon potential are taken from Ref.\,\cite{Wang2017}.  For the g.s. of $^6$Be, the depth $V_0$ has been readjusted to reproduce the experimental decay energy \cite{ENSDF,Webb2019} ($Q_{2p}$ = 1.372\,MeV). Since the  g.s. of $^6$He is bound with respect to $2n$ decay, to study the properties of $2n$ decays  at long times/distances, by readjusting $V_0$ we artificially created an unbound g.s. of $^6$He   with $Q_{2n}$ = 1\,MeV, referred to as $^6$He$^\prime$ in the following. The calculated two-nucleon decay widths are 64\,keV and 241\,keV for $^6$Be and $^6$He$^\prime$, respectively.

To benchmark our time-dependent approach, we  carried out time propagation using  Green's function  $\hat{G} = (E- \hat{H} +i\eta)^{-1}$. The corresponding time evolution operator can be written as the Fourier transform of $\hat{G}$:
\begin{equation}\label{Green_function}
	e^{-i \frac{\hat{H}}{\hbar} t}=\frac{e^{\frac{\eta}{\hbar} t}}{2 \pi i} \mathcal{F}\left(\hat{G} , E \rightarrow \frac{t}{2 \pi \hbar}\right).
\end{equation}
We adopt the Berggren basis expansion with the ``off-diagonal method'' \cite{Michel2011} to avoid the singularities stemming from the denominator of Green's function and the integration of Coulomb interaction in the asymptotic region \cite{Combescot2017,Hlophe2019}.

\begin{figure}[!htb]
\includegraphics[width=0.8\columnwidth]{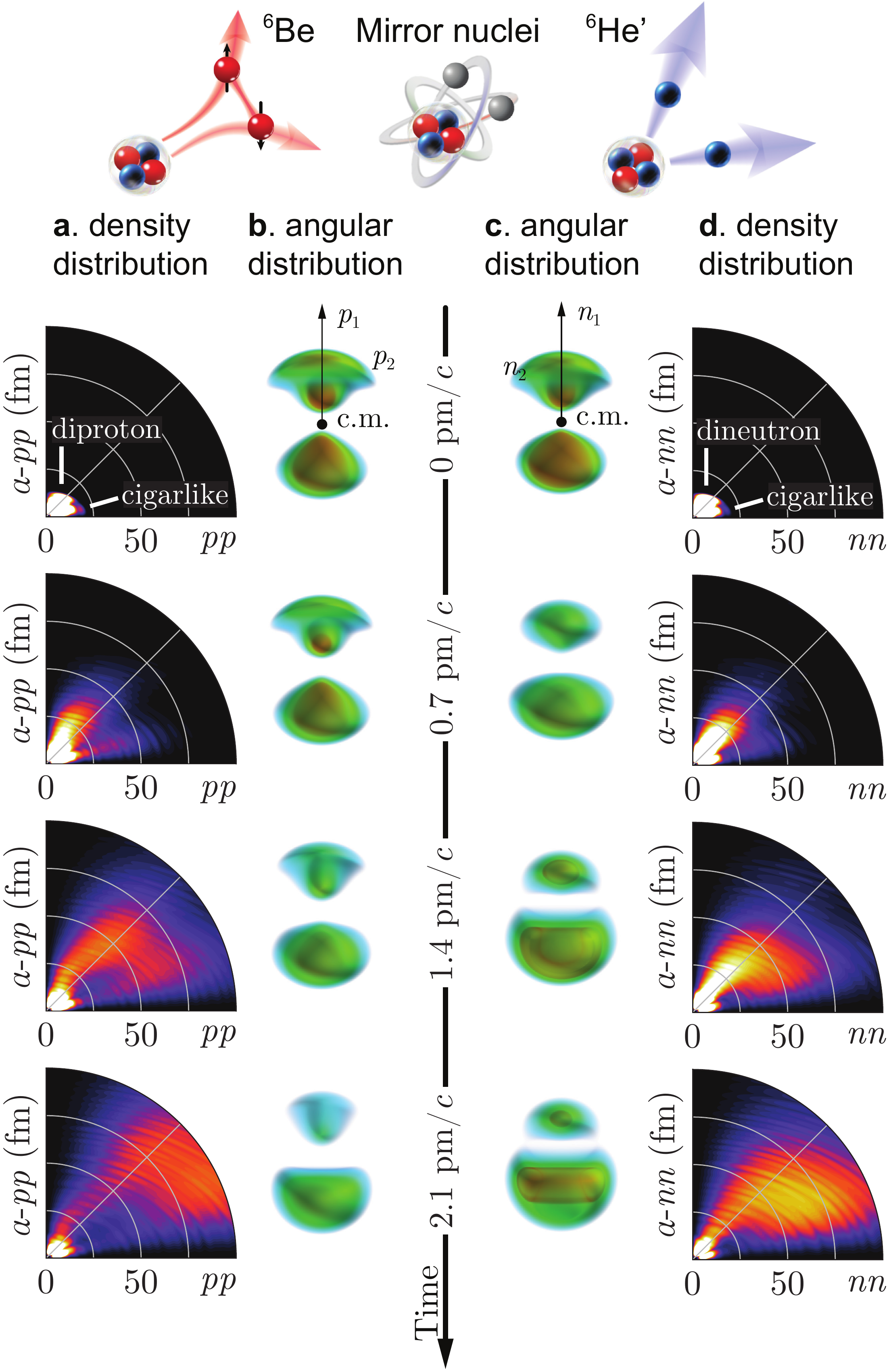}
\caption{The density and momentum distributions of  two-nucleon decays from the  g.s. of ${^6}$Be (left) and ${^6}$He$^\prime$ (right) for four different time slices. The density distributions are shown in the Jacobi-T coordinates (see Fig.\,\ref{Jacobi_coordinates} and SM \cite{SM}). The momentum distribution of the second nucleon  is shown with respect to the momentum of the first nucleon. To show the asymptotic wave function clearly, all the particle densities (in fm$^{-1}$) are multiplied by the polar Jacobi coordinate $\rho$. The dimensionless momentum (angular) distributions are divided by the total momentum $k$.}\label{Density_evolution}
\end{figure}

{\it Different dynamics of ${2p}$ and ${2n}$ decays}.---For the light $2p$ emitters,  both direct and sequential decays are possible \cite{Bochkarev1989,Pfutzner2012}. The decay of $^6$Be represents one of such cases \cite{Papka2010}. Here, the neighboring g.s. of $^5$Li is a broad resonance with a proton decay width of $\Gamma=1.23$\,MeV. The three-body decay of $^6$Be is not completely understood \cite{Pfutzner2012,Barker2003,Blank2008,Alvarez2008,Grigorenko2009,Grigorenko2009_2,Egorova2012,Chudoba2018}. Indeed, the diproton structure predicted by theory and the measured broad angular correlation of the emitted protons have been viewed as self-contradictory. Figure~\ref{Density_evolution}a,b shows the calculated evolution of the $2p$ density and momentum distribution for the g.s. of $^6$Be  over a broad time range. At the beginning ($t=0$) when  the wave function is fairly localized inside the nucleus, the density distribution shows two maxima for $^6$Be associated with diproton/cigarlike configuration characterized by small/large relative distance between valence protons.  (For a better visualization of the initial $2p$ wave function, see Ref.\,\cite{Wang2019}.)  
During the early stage of the decay, two strong flux branches dominate, see Fig.~\ref{Density_evolution}b. The primary branch corresponds to the protons being emitted at small opening angles, which indicates that a diproton structure is present  during the tunneling phase. This can be understood in terms of the nucleonic pairing, which favors low angular momentum amplitudes and  hence lowers the centrifugal barrier and increases the $2p$ tunneling probability  \cite{Grigorenko2009_2,Oishi2014,Oishi2017,Wang2019}. The secondary branch corresponds to protons  emitted in opposite directions. While they are spatially apart, these protons seem to be correlated and decay simultaneously according to their similar proton-core distances; this nicely reveals a three-body nature of the process.

After tunneling through the Coulomb barrier, the two emitted protons tend to gradually separate due to  the Coulomb repulsion. This is reflected in the bent trajectory of the diproton decay branch and the gradual reduction of the momentum alignment seen in Fig.~\ref{Density_evolution}a,b. Eventually, the $2p$ density becomes spatially diffuse, which is consistent with the broad angular distribution measured in Ref.\,\cite{Egorova2012}. One may notice that even beyond 100\,fm (at $t \approx$ 2\,pm/$c$), the Coulomb repulsion tends to reduce the inter-proton correlation. According to our calculations, the angular correlation  starts to stabilize only after very long times greater than 9\,pm/$c$. Therefore, in order to make meaningful estimates of asymptotic observables, very long propagation times are indeed required.

\begin{figure}
\includegraphics[width=1.0\columnwidth]{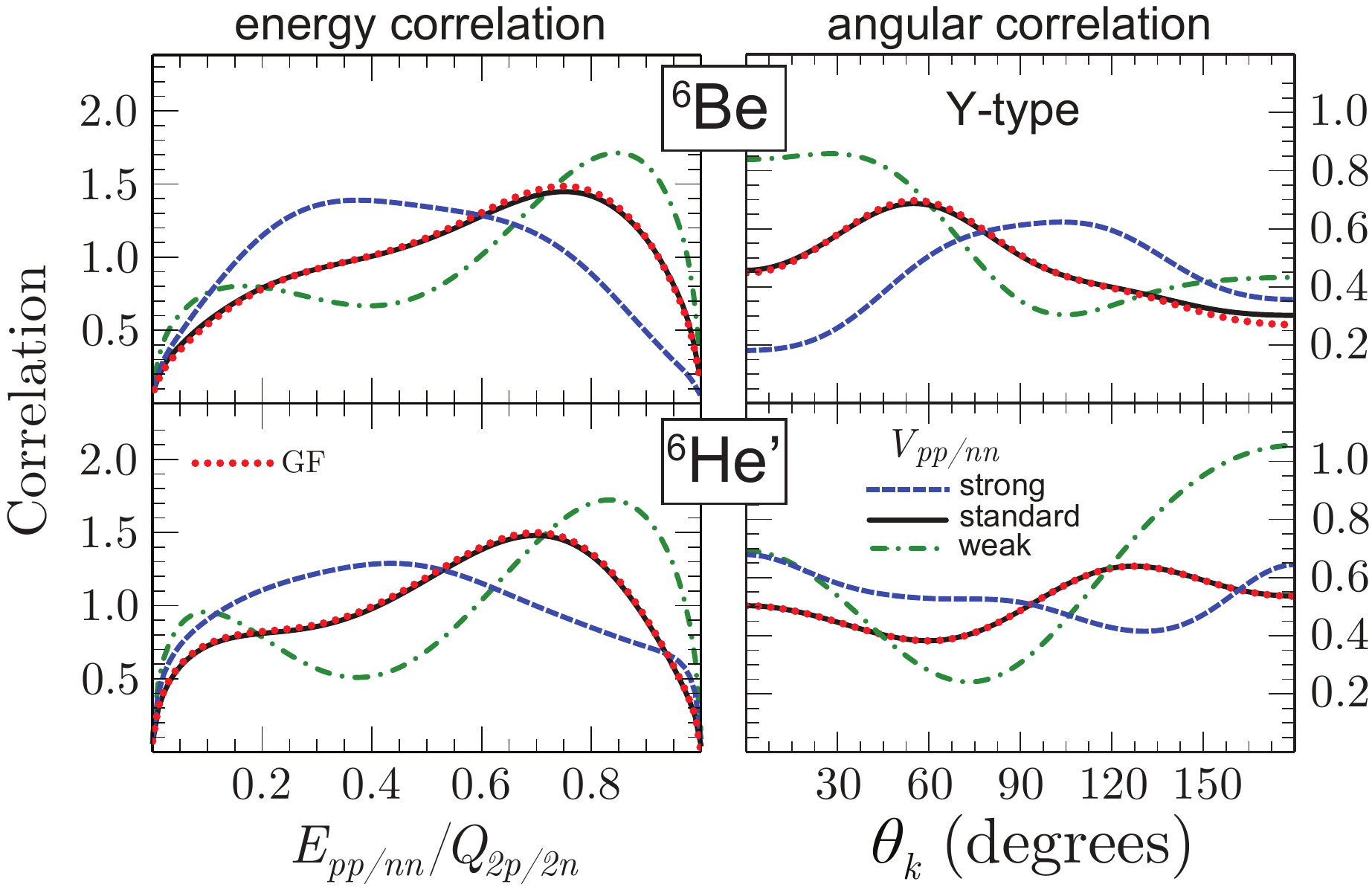}
\caption{
Asymptotic energy (left) and angular (right) correlations   of emitted nucleons  from the g.s. of $^6$Be (top) and $^6$He$^\prime$ (bottom) calculated at ${t=15}$\,pm/${c}$ with different strengths of  the Minnesota interaction \cite{Thompson1977}: 
standard (solid line), strong (increased by 50\%; dashed line),  and weak (decreased by 50\%; dash-dotted line). Also shown are the benchmarking results obtained within Green's function method (GF; dotted line) using the standard  interaction strength.
$\theta_k$ is the opening angle between $\vec{k}_x$ and $\vec{k}_1$ in the Jacobi-Y coordinate system, 
 and $E_{pp/nn}$ is the kinetic energy of the relative motion of the emitted nucleons (see Fig.\,\ref{Jacobi_coordinates} for definitions).
}\label{Correlation_A6}
\end{figure}

The mirror partner of  $^6$Be is the $2n$ halo system $^6$He. To study the nuclear-Coulomb interplay in the two-nucleon decay, we study the artificially-unbound   $^6$He$^\prime$. As seen in Fig.~\ref{Density_evolution},  at small times the density  of $^6$He$^\prime$ looks similar to that of $^6$Be due to the isospin symmetry of nuclear force. However, since the Coulomb repulsion is absent in the $^6$He$^\prime$ case,  the dineutron decay branch is more pronounced as the emitted neutrons keep maintaining their spatial correlations in time. As a result, the asymptotic nucleon-nucleon correlations displayed in Fig.\,\ref{Correlation_A6} are quite different between $^6$Be and $^6$He$^\prime$. 

To illustrate our benchmarking procedure, Fig.\,\ref{Correlation_A6} shows the comparison between our time-dependent approach  and the Green's function method. Both techniques
produce results that  are practically identical.

\begin{figure}[!htb]
\includegraphics[width=0.9\columnwidth]{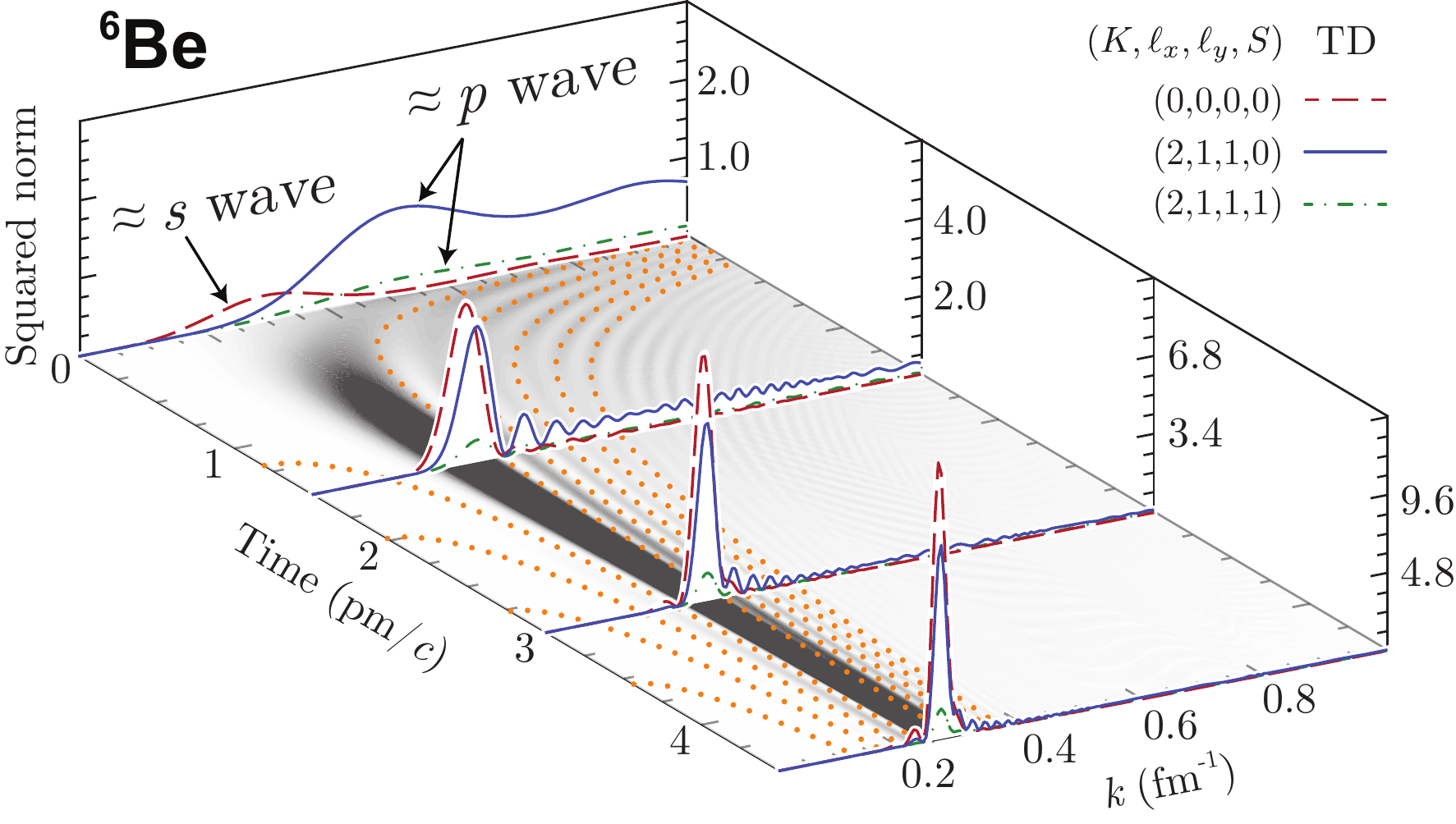}
\caption{Time evolution of the wave functions of ${^6}$Be. Configurations are labeled as $(K,\ell_x,\ell_y,S)$ in Jacobi-T coordinates. The projected contour map represents the sum of all the configurations in momentum space; the interference frequencies are marked by dotted lines corresponding to different ${\mathfrak n}$-values. }\label{Configuration_evolution}
\end{figure}

{\it Impact of nucleon-nucleon interaction on correlations between emitted nucleons.}--
To gain more insight into the interplay between the initial-state and final-state interactions, we studied the decay properties of $^{6}$Be and $^{6}$He$^\prime$ as a function of the nucleon-nucleon Minnesota interaction strength $V_{pp/nn}$. As shown in Fig.\,\ref{Correlation_A6} and SM \cite{SM}, the attractive nuclear force is not only responsible for the presence of  correlated dinucleons in the initial state, but it also significantly impacts asymptotic energy correlations and angular correlations in the Jacobi-Y angle $\theta_k$. This indicates that, even though the initial-state  correlations are largely lost  in the final state, some fingerprints of the dinucleon structure can still manifest themselves in  the asymptotic observables. Interestingly, the asymptotic angular correlations  in the Jacobi-T angle $\theta_k'$ hardly depend on $V_{pp/nn}$ \cite{SM}. This suggests that this observable  is not particularly  useful when assessing nucleon-nucleon correlations in the initial state.

By comparing correlation results for  $^{6}$Be and $^{6}$He$^\prime$ in Fig.\,\ref{Correlation_A6} one can assess the role of Coulomb interaction in the $2p$ decay. One can see that the energy correlations are rather similar for both  nuclei. This is certainly not the case for the angular correlations: their patterns are markedly different, independent of the interaction strength $V_{pp/nn}$.

{\it Evolution of two-nucleon wave function}.--Figure\,\ref{Configuration_evolution} illustrates the propagation of the wave function of $^6$Be. At $t = 0$, the g.s wave function shows a broad momentum distribution that is indicative of spatial localization. At long times, the emitted nucleons move  with the well-defined total momentum shown by a narrow resonance peak corresponding to the resonance's energy \cite{IdBetan2018}. The gradual transition from the broad to narrow momentum distribution exhibits a pronounced interference pattern, which is universal for two-nucleon decays. The interference frequencies, shown by dotted lines in Fig.~\ref{Configuration_evolution}, can be approximated by $(\frac{\hbar^2}{2m} k^2 - Q_{2p/2n}) t = {\mathfrak n}\pi \hbar$, where ${\mathfrak n}$ = 1, 3, 5 $\cdots$, i.e., they explicitly depend on the $Q_{2p/2n}$ energy  (see SM \cite{SM} for details).

Interestingly, the configuration evolution also reveals a unique feature of three-body decay. As seen in Fig.\,\ref{Configuration_evolution}, the initial g.s. of $^{6}$Be is dominated by the $p$-wave ($\ell = 1$) components, and the small $s$-wave ($\ell = 0$) component comes from the non-resonant continuum. As the system evolves, the weight of the $s$-wave component -- approximately corresponding to the Jacobi T-coordinate configuration $(K,\ell_x,\ell_y,S)$ = (0,0,0,0) -- gradually increases and eventually dominates because it experiences no centrifugal barrier. Such a behavior can never happen in the single-nucleon decay due to the conservation of orbital angular momentum but is present for the two-nucleon decay as 
 a correlated di-nucleon involves components with different $\ell$-values \cite{Catara1984,Pillet2007,Hagino2014,Fossez2017}.  Moreover,  for $2p$ decays  the Coulomb potential and kinetic energy do not commute  in the asymptotic region \cite{Grigorenko2009} and this results
 in additional configuration mixing.

{\it Summary}.---To study the mechanism  of two-nucleon decay, theoretical models must fully control the behavior of the decaying system at large distances and long propagation times. To this end, we developed a realistic time-dependent framework that allows for precise three-body solutions asymptotically. 
Our calculations demonstrate the different dynamics of $2p$ and $2n$ decays. The  initial-state $pp$ correlations are largely lost due to the Coulomb repulsion between escaping protons.  
We showed that the energy and angular correlations in  the Jacobi-Y angle $\theta_k$ between emitted nucleons strongly depend on the  initial-state structure.

{\it Acknowledgements}.---Discussions with Robert Charity, Marek P{\l}oszajczak, Jimmy Rotureau, and Lee Sobotka are gratefully acknowledged. We appreciate helpful comments from Samuel Giuliani and Terri Poxon-Pearson. This material is based upon work supported by the U.S.\ Department of Energy, Office of Science, Office of Nuclear Physics under award numbers DE-SC0013365 (Michigan State University) and DE-SC0018083 (NUCLEI SciDAC-4 collaboration).

\bibliography{references}

\end{document}